\documentclass[twocolumn,preprintnumbers,amsmath,amssymb,aps,prl,floatfix]{revtex4}

\usepackage{graphicx}
\usepackage{dcolumn}
\usepackage{bm}

\def\apj #1 #2 #3 {#1, ApJ, {\bf #2}, #3}
\def\apjl #1 #2 #3 {#1, ApJ, {\bf #2}, L#3}
\def\apjs #1 #2 #3 {#1, ApJS, {\bf #2}, #3}
\def\aap  #1 #2 #3 {#1, A\&A, {\bf #2}, #3}
\def\mnras #1 #2 #3 {#1, MNRAS, {\bf #2}, #3}
\def\pra #1 #2 #3 {#1, Phys.~Rev.~A., {\bf #2}, #3}
\def\prb #1 #2 #3 {#1, Phys.~Rev.~B., {\bf #2}, #3}
\def\prc #1 #2 #3 {#1, Phys.~Rev.~C., {\bf #2}, #3}
\def\prd #1 #2 #3 {#1, Phys.~Rev.~D., {\bf #2}, #3}
\def\pre #1 #2 #3 {#1, Phys.~Rev.~E., {\bf #2}, #3}
\def\prl #1 #2 #3 {#1, Phys.~Rev.~Lett., {\bf #2}, #3}
\def\plb #1 #2 #3 {#1, Phys.~Lett.~B., {\bf #2}, #3}
\def\science #1 #2 #3 {#1, Science., {\bf #2}, #3}
\def\nature #1 #2 #3 {#1, Nature., {\bf #2}, #3}
\def\nphysa #1 #2 #3 {#1, Nucl.~Phys.~A., {\bf #2}, #3}
\def\nphysb #1 #2 #3 {#1, Nucl.~Phys.~B., {\bf #2}, #3}
\def\nphysbs #1 #2 #3 {#1, Nucl.~Phys.~B.~Suppl., {\bf #2}, #3}

\def\h#1{\hbox{${}^{#1}$H}}

\def\h502{\hbox{$ h^{2}_{50}$}}

\def\la{\mathrel{\mathpalette\fun <}}

\def\fun#1#2{\lower3.6pt\vbox{\baselineskip0pt\lineskip.9pt
  \ialign{$\mathsurround=0pt#1\hfil##\hfil$\crcr#2\crcr\sim\crcr}}}

\begin{document}

\preprint{}

\title{WMAP Constraints on Decaying Cold Dark Matter}

\author{Kiyotomo Ichiki$^{1,2}$, Masamune Oguri$^3$ and Keitaro Takahashi$^3$}
\affiliation{
$^1$National Astronomical Observatory, 2-21-1, Osawa, Mitaka, Tokyo
181-8588, Japan \\
$^2$Department of Astronomy, University of Tokyo, Tokyo 113-0033, Japan\\
$^3$Department of Physics, University of Tokyo, Tokyo 113-0033, Japan 
}
\date{\today}

\begin{abstract}
 We re-formulate cosmological perturbations in the decaying cold dark
 matter model, and calculate cosmological microwave background (CMB)
 anisotropies. By comparing our theoretical predictions with recent
 observational data from the Wilkinson Microwave Anisotropy Probe
 (WMAP), we derive a new bound on the abundance and lifetime of decaying 
 dark matter particles. We show that the data of WMAP alone do not
 prefer the decaying cold dark matter model:  the lifetime is
 constrained to $\Gamma^{-1} \ge 123$ Gyr at $68\%$ C.L. ($52$ Gyr at
 95.4$\%$ C.L.) when cold dark matter consists only of such decaying
 particles. We also consider a more general case in which cold dark
 matter consists of both stable and decaying particles, and show that
 the constraint generalizes down to 
  $\Omega_{\rm DDM}h^2 \la -0.5(\Gamma^{-1}/1{\rm
  Gyr})^{-1}+0.12$ for $\Gamma^{-1} \ge 5\mbox{Gyr}$
 at $95.4\%$ C.L.. These bounds are robust and widely applicable, because
 they are derived only from gravitational effects and therefore do not
 rely on the details of decay channels or decay products.
\end{abstract}

\maketitle

{\it Introduction.---}
The existence of cold dark matter (CDM) is now widely accepted from
numerous kinds of astronomical phenomena, such as rotation curves
in galaxies, anisotropies in cosmic microwave background (CMB), and
X-ray emitting clusters of galaxies combined with the low cosmic baryon
density predicted by big bang nucleosynthesis.
The nature of CDM, however, is still one of the biggest mysteries in
cosmology. Indeed, some discrepancies on galactic and sub-galactic
scales in standard CDM cosmology have stimulated numerous proposals to
modify the standard CDM model \cite{ostriker03}.

The decaying cold dark matter (DCDM) model is one of such proposals to
solve discrepancies in the standard CDM model. For instance, Cen proposed
the DCDM model to solve both over-concentration problem of the dark
matter halos and over-production problem of small dwarf galaxies
\cite{cen01}. The authors also showed in previous papers
\cite{oguri03} that introducing DCDM can improve the fits of
observational data sets of Type Ia SN, mass-to-light ratios and X-ray
gas fraction of clusters of galaxies, and evolution of cluster abundance. 
The DCDM model may be also useful to solve other important problems such 
as ultra-high energy cosmic rays above the Greisen-Zatsepin-Kuzmin
cut-off of the spectrum \cite{kuzmin98} and the early reionization of
the universe \cite{kasuya03}.  

Although theoretical candidates for DCDM have been proposed, their
predictions for lifetime of decaying particles cover a large range
of values, from $10^{-2}$ Gyr to $10^{11}$ Gyr
\cite{Chung98}.  This indicates that it is quite
important to constrain (possible) lifetime of dark matter particles from
astronomical observations. One popular way to constrain the lifetime 
is to use decay products, e.g., diffuse gamma ray background
observations \cite{Dolgov:1981hv}. However, realistic simulation which
takes all energy dissipation processes into account showed that even the
particles with lifetime as short as a few times of the age of the
universe still are not ruled out by recent observations
\cite{Ziaeepour:2000rc}.
Moreover, the results are dependent on the
assumed decay channel while we do not know what dark matter is.

In this paper, we derive cosmological constraints on the abundance and
lifetime of decaying dark matter particles from CMB anisotropies. 
We re-formulate cosmological perturbations in the decaying cold dark
matter model and compare them with accurate measurements by the
Wilkinson Microwave Anisotropy Probe (WMAP)
\cite{Spergel:2003cb}.
Here we assume only that dark matter particles decay into relativistic
particles, whatever they may be. Thus our constraints are different from
those derived from decay products such as diffuse gamma ray background
observations which need to assume the details of the decay products or
decay channels. Throughout this paper we concentrate our attention on a
flat universe. 

{\it DCDM cosmology: Basic Equations.---}
To make theoretical predictions of CMB anisotropies in the DCDM model, 
first we must formulate cosmological perturbation theory. Similar works
can be found in references \cite{Kaplinghat:1999xy, Bharadwaj:1997dz} in 
which they discussed decaying neutrinos. The present work differs from
them in that we consider decays of cold dark matter, and we treat both
the abundance and the lifetime of decaying particle as free parameters.

The equations of background energy densities for DCDM particles (DDM)
and its relativistic daughter particles (DR) are given by
\begin{eqnarray}
 \dot{\rho}_{\rm DDM}&=&-3{H}\rho_{\rm DDM} -a \Gamma \rho_{\rm DDM}~, \\
 \dot{\rho}_{\rm DR}&=&-4{H}\rho_{\rm DR} +a \Gamma \rho_{\rm DDM}~.
\end{eqnarray}
Here dot denotes conformal time derivative and $\Gamma$ is the decay
width of the dark matter. 
The equation of state parameters $w = P/\rho$ are $w_{\rm DDM}=0~$, 
and $w_{\rm DR}=1/3~$, respectively. 
We also define effective equation of state,
\begin{equation}
  w_{\rm eff}^{\rm DDM}\equiv\frac{a\Gamma}{3H}~,\quad w_{\rm eff}^{\rm DR}
\equiv \frac{1}{3}-w_{\rm eff}^{\rm DDM}\left(\frac{\rho_{\rm DDM}}{\rho_{\rm DR}}\right)~,
\end{equation}
which are defined by the evolution of energy density,
$\dot\rho_i=-3{H}(1+w_{\rm eff})\rho_i$.

Now let us turn to the cosmological perturbations.
In the conformal Newtonian gauge, the line element is given by
\begin{equation}
 ds^2=a(\tau)^2\left[-(1+2\psi) d\tau^2 + (1+2\Phi)\delta_{ij}dx^i
 dx^j\right]~.
\end{equation}
where $a(\tau)$ is cosmic scale factor, $\psi$ and $\Phi$ are 
perturbations around flat Friedmann-Robertson-Walker metric.

To derive the equation of DCDM we start from the Boltzmann equation for
non-relativistic matter with decaying term and 
take its zeroth and first moments by momentum integration:
\begin{eqnarray}
\dot n+\partial_i (n v^i)+3(H+\dot \Phi)n = -a\Gamma n~, \\
\partial_\tau(nv^j)+4Hnv^j+n\partial_j \psi=-a\Gamma nv^j~.
\end{eqnarray}
We decompose $n(\tau,x^i)=n_0(\tau)(1+\delta(\tau,x^i))$ and use
gauge invariant density perturbations defined by $D^g=\delta+3(1+w_{\rm
eff})\Phi$. Then we have the following equations for DCDM: 
\begin{eqnarray}
\dot D^g_{\rm DDM} &=& -kv_{\rm DDM}+3(\dot w_{\rm eff}^{\rm DDM}\Phi +
w_{\rm eff}^{\rm DDM}\dot\Phi)~, \\
\dot v_{\rm DDM}&=&-Hv_{\rm DDM}+k\psi~,
\end{eqnarray}
here we have defined $v\equiv i(k_j/k)v^j$.

Once we have equations for DCDM we have those for its daughter
radiation (DR) from the consequence of energy-momentum conservation:
\begin{eqnarray}
 \dot D_{\rm DR}^g &=& -\frac{4}{3}kv_{\rm DR}+(3w_{\rm eff}^{\rm DR}-1)(\dot \Phi + HD_{\rm DR}^g) \nonumber \\
&+&3\dot w_{\rm eff}^{\rm DR}\Phi
+3(1-2w_{\rm eff}^{\rm DR}-3w_{\rm eff}^{\rm DR^2})H\Phi \nonumber \\
&+&a\Gamma\left(\frac{\rho_{\rm DM}}{\rho_{\rm DR}}\right)\left(D_{\rm DDM}^g-3(1+w_{\rm eff}^{\rm DDM})\Phi\right) \label{continuity_for_DR}~,\\ 
 \dot v_{\rm DR}&=&(3w^{\rm DR}_{\rm eff}-1){H}v_{\rm DR}+k\psi-\frac{1}{6}k\Pi_{\rm DR}+\frac{k}{4}D_{\rm DR}^g \nonumber \\
&-&\frac{3}{4}(1+w^{\rm DR}_{\rm eff})k\Phi+ a\Gamma v_{\rm DDM} \left(\frac{3\rho_{\rm DDM}}{4\rho_{\rm DR}}\right)~. 
 \label{euler_for_DR}
\end{eqnarray}

Further, we need Boltzmann hierarchy for $l \ge 2$ moments of daughter
radiation, where $l$ stands for multipole moment in a Legendre expansion
of perturbed distribution function. 
Again we begin with the Boltzmann equation for relativistic particle
\cite{Ma:1995ey},
\begin{eqnarray}
 f_0^{\rm DR}\frac{\partial \Psi^{\rm DR}}{\partial \tau}&+&\frac{\partial
 f_0^{\rm DR}}{\partial \tau}\Psi^{\rm DR}+i(\vec{k}\cdot
 \hat{n})\Psi^{\rm DR} \nonumber \\
&+&\frac{d f_0^{\rm DR}}{d q}(q\dot\Phi-iq(\vec{k}\cdot\hat{n})\psi)=\left(\frac{\partial f}{\partial \tau}\right)_{\rm c}~,
\end{eqnarray}
where $\vec{q}=q\hat{n}$ is the comoving 3-momentum with $n^i n_i = 1$.
We wrote the distribution function of daughter radiation with background
distribution and perturbation around it as
 $f^{\rm DR}(x^i,q,n_j,\tau)=f_0^{\rm DR}(q,\tau)(1+\Psi^{\rm DR}(x^i,q,n_j,\tau))$.
We should note that, unlike the standard CDM, $f_0^{\rm DR}$ is time
{\it dependent} in the DCDM model. To describe the decay process, let us
consider the collision term:
\begin{equation}
 \left(\frac{\partial f}{\partial \tau}\right)_{\rm c}=a\Gamma \frac{m_{\rm DDM}}{q}f_0^{\rm DDM}(1+\Psi^{\rm DDM})~,
\end{equation}
where $f^{\rm DDM}=f_0^{\rm DDM}(1+\Psi^{\rm DDM})$ and $m_{\rm DDM}$
are the distribution function and mass of DCDM particles, respectively.
Then we obtain
\begin{eqnarray}
 \frac{\partial}{\partial \tau}F^{\rm DR}+ik\mu
  F^{\rm DR}+4(\dot\Phi+ik\mu\psi)\nonumber &&\\ 
&&\hspace*{-55mm}=-a\Gamma
  \frac{\rho_{\rm DDM}}{\rho_{\rm DR}}F^{\rm DR}+\frac{a\Gamma m_{\rm DDM}\int q^2 dq~f_0^{\rm DDM}\Psi^{\rm DDM}}{(a^4\rho_{\rm DR})/(4\pi)}~,
\label{1st_order_intBoltz}
\end{eqnarray}
where $\mu \equiv \vec{k}\cdot \hat{n}$, and we defined
$ 
F^{\rm DR}(\vec{k},\hat{n},\tau)\equiv \int q^2 dq~q f_0^{\rm
DR}\Psi^{\rm DR}/{\int q^2 dq~q f_0^{\rm DR}}
~.
$
The two terms in r.h.s of
Eq.(\ref{1st_order_intBoltz}) are new ones in the DCDM model. 
The first term comes from the fact that evolution of background energy
density of daughter radiation is different from decoupled radiations
which diminish as $a^{-4}$.
The second term corresponds to the flow from the dark matter to daughter
radiation in first order perturbation. This is complicated to compute in
general \cite{Kaplinghat:1999xy}, but it can
be easily described in a fluid approximation (see Eq. (\ref{continuity_for_DR}) and
Eq.(\ref{euler_for_DR}) for $l=0$ and $l=1$, respectively).
For higher multipoles ($l \ge 2$), this term vanishes since the
perturbation in the dark matter ($\Psi^{\rm DDM}$) does have only $l=0$
and $1$ perturbations which correspond to density and velocity
perturbations, respectively. Finally, we have hierarchy for daughter
radiation $(l \ge 2)$, 
\begin{equation}
\dot M_l =  k \left( \frac{l}{2l-1} M_{l-1} 
 - \frac{l+1}{2l +3} M_{l+1}\right)
 - a\Gamma\left(\frac{\rho_{\rm DDM}}{\rho_{\rm DR}}\right)M_l~,
\end{equation}
where $M_l$ is the coefficient in a Legendre expansion,
 $\frac{1}{4}F^{\rm DR}+\Psi \equiv M=\sum (-i)^l M_l(\tau,\vec{k}) P_l(\mu)~$.
These coefficients are related with above perturbation variables by
$D^g_{\rm DR} = 4 M_0$, $v_{\rm DR}=M_1$, and
$\Pi_{\rm DR}=(12/5)M_2$. 

{\it CMB constraints and Discussions.---}
We are now ready to compute theoretical CMB spectrum in DCDM cosmology.
We first consider a simple DCDM cosmology in which CDM 
component consists only of decaying particles, and constrain the
abundance and lifetime of DCDM particles from observed CMB anisotropies
of the WMAP. Next, as a generalization of DCDM cosmology, we investigate the
universe in which CDM consists both of stable (standard) and decaying
particles.

The likelihood functions we calculate are given by 
Verde et al.\cite{Verde:2003ey}.
To include the decay of dark matter particles described above, we
calculate theoretical angular power spectrum of CMB fluctuation $(l
(l+1)C_l/(2\pi))$ using a modified Boltzmann code of CAMB \cite{camb},
which is based on a line of sight integration approach \cite{cmbfast}.

 \begin{figure}
 \includegraphics[width=0.45\textwidth]
{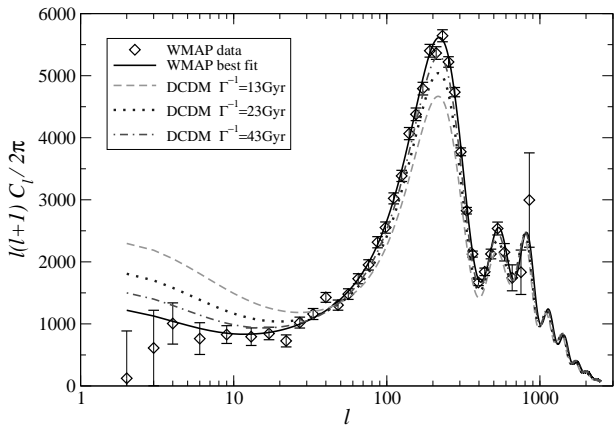}
  \caption{CMB angular power spectrum with and without decay of CDM
  particles, and observational data from the WMAP. The solid line 
  corresponds to the best-fit standard CDM model. The other
  lines are DCDM models with $\Gamma^{-1}=13$ ({\it
  dash}), $23$ ({\it dotted}), and $43$ Gyr ({\it dash-dotted}).
  For each lines, all cosmological parameters (except for
  lifetime) are fixed to WMAP optimal values to demonstrate how decays
  of dark matter particles modify the angular power spectrum.} 
\label{fig:L1}
 \end{figure}

An illustration of CMB power spectrum in the DCDM models is shown in
Fig. \ref{fig:L1}. The decay of CDM particles modify the CMB spectrum
mainly in three ways. First, the modified evolution of dark matter
changes the expansion history of the universe and generally causes the
shorter look back time to the photon last scattering surfaces. This
pushes the acoustic peaks in the CMB spectrum toward the larger scales
(smaller $l$). Second, decays of dark matter particles mean that there
is more dark matter at earlier times. This results in smaller anisotropies
around the first acoustic peak. Third, decays of gravitational potential
at later epochs due to decays of dark matter particles lead to a larger
late integrated Sachs-Wolfe (LISW) effect. The LISW significantly
enhances CMB anisotropies at low multipoles. Thus, the amplitudes and
locations of the peaks in the power spectrum of microwave background
fluctuations \cite{hu} can in principle be used to constrain the DCDM
model. 

\begin{figure}
   \rotatebox{0}{\includegraphics[width=0.4\textwidth]{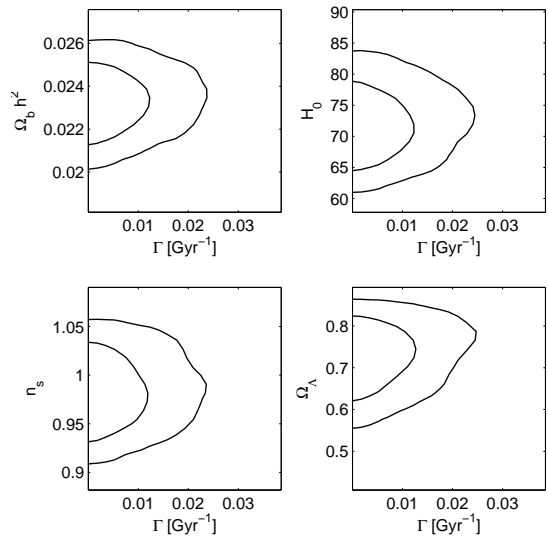}}
 \caption{Contours of constant relative probabilities in two dimensional
 parameter planes. Lines correspond to $68\%$ and $95.4\%$ confidence 
 limits. }
 \label{fig:L2}
\end{figure}

It is well known, however, that there are other cosmological parameters
which also modify the CMB spectrum. Therefore, we have to generate the
full probability distribution function and marginalize over nuisance
parameters to obtain the constraint on parameter(s) which we are
interested in. To realize this, we followed the Markov Chain Monte Carlo
approach \cite{MCMC} and explore the likelihood in seven dimensional
parameter space i.e., five standard parameters, $\Omega_{\rm b} h^2$ (baryon
density), $h$ (Hubble parameter), $z_{\rm re}$ (reionization redshift),
$n_{\rm s}$ (power spectrum index), $A_{\rm s}$ (overall amplitude), and
two DCDM parameters, $\Omega_{\rm DDM} h^2$ (DCDM density) and $\Gamma$
(decay width). 

\begin{figure}
   \rotatebox{-90}{\includegraphics[width=0.35\textwidth]{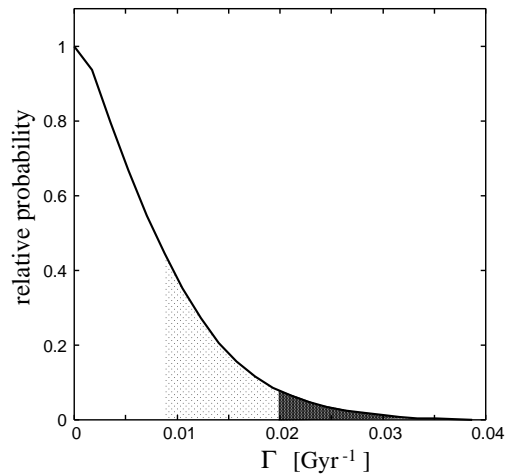}}
\caption{Marginalized probability distribution of the decay parameter
 $\Gamma$ in Gyr$^{-1}$. Confidence limits of $68\%$ and $95.4\%$ are shown by
 shaded regions.}
 \label{fig:L3}
\end{figure}

Our results are shown in Figs. \ref{fig:L2} and \ref{fig:L3}.
An interesting point is that the parameter of the DCDM model, $\Gamma$,
dose not degenerate with the other cosmological parameters very much as
one can see in Fig. \ref{fig:L2}. This means that the change in the CMB
spectrum from $\Gamma$ cannot be mimicked by other standard parameters.
This is the reason why CMB can put strong constraint on the lifetime of
DCDM particles.  Figure  \ref{fig:L3} shows marginalized likelihood of
the lifetime of DCDM particles:  the constraint is $\Gamma^{-1} \ge 52$
Gyr at $95.4\%$ C.L., and $\Gamma^{-1} \ge 123$ Gyr at $68\%$ C.L.. We do 
not find any signal to prefer the DCDM model to standard CDM cosmology. 

\begin{figure}
   \rotatebox{-90}{\includegraphics[width=0.2\textwidth]{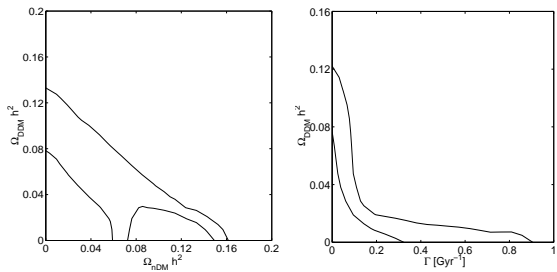}}
\caption{The same as Fig.\ref{fig:L2}, but for cosmology in which dark
 matter consists of two species (stable and decaying dark matters).
 The densities of stable and decaying dark matters are denoted by 
 $\Omega_{\rm nDM}$ and $\Omega_{\rm DDM}$, respectively. }
 \label{fig:L4}
\end{figure}

To obtain more general constraints on the DCDM models, next we consider
a cosmological model in which CDM consists of two types of particles,
i.e., one is stable and the other is unstable. We performed the same CMB
likelihood analysis described above with another cosmological parameter,
$\Omega_{\rm nDM}$, which stands for the current density of stable dark
matter in critical density units. Our results are shown in Fig.
\ref{fig:L4}. The results are easily understood: first, the universe
should have either CDM or DCDM component $\Omega_{\rm CDM}h^2 \approx
0.1$, though CDM are slightly favored compared with DCDM, (left of Fig. \ref{fig:L4}); second, if the
dark matter mainly consists of unstable particle, its lifetime is tightly
constrained (right of Fig. \ref{fig:L4}) as shown in Figs.
\ref{fig:L2} and \ref{fig:L3}. However, as the fraction of decaying
components becomes smaller, constraints on $\Gamma^{-1}$ becomes weaker.
We find that the lifetime of unstable CDM particle is roughly
constrained to
\begin{displaymath}
 \left.
 \begin{array}{ll}
  \Omega_{\rm DDM}h^2 \la -0.5\left(\frac{\Gamma^{-1}}{1{\rm
  Gyr}}\right)^{-1}+0.12 &\;\;\mbox{for}\ \Gamma^{-1} \ge 5\mbox{Gyr}\\
  \Omega_{\rm DDM}h^2 \la -0.03\left(\frac{\Gamma^{-1}}{1{\rm
  Gyr}}\right)^{-1}+0.026 &\;\;\mbox{for}\ \Gamma^{-1} \le 5\mbox{Gyr}\\
 \end{array}
\right.
\end{displaymath}
at $95.4\%$ C.L.. 
Bounds on DCDM would not be improved so much by future
experiments since DCDM affects CMB fluctuation only in rather large scales
(see Fig. \ref{fig:L1}), and errors in these scales are already limited
by cosmic variance. In this sense our results are robust and the tightest
constraints on DCDM cosmology from CMB anisotropies.

Before conclusion, we briefly mention the reionization
history in DCDM models. If the energy of decaying particles is
transformed into high energy photons, they may become a source of
ionizing photons and lead an early reionization of the universe.
This results in the suppression of acoustic peaks in the CMB 
spectrum\cite{Bean:2003kd}, which is similar to the purely gravitational
effect considered here. Therefore, constraints on the decay rate
might be stronger when we include the ionizing photons produced from the
decays. However, this process depends  on the details
of the decay process, such as the power index of the injected spectrum
and decay modes to a certain extent.  On the other hand, our constraints are based only on
the minimal assumption that dark matter particles decay into
relativistic particles. Thus our result is the most conservative one.

{\it Conclusion.--}
We showed that even the current CMB data alone put strong constraint on
the lifetime of cold dark matter to $\Gamma^{-1} \ge 123$ Gyr at $68\%$
C.L. ($52$ Gyr at $95.4\%$). 
By considering both stable and decaying dark
matters, the constraint on DCDM with lifetime larger than $5$ Gyr
generalizes down to
$\Omega_{\rm DDM}h^2 \la -0.5(\Gamma^{-1}/1{\rm Gyr})^{-1}+0.12$
at $95.4\%$ C.L..
The results are based on the minimal assumption, free from 
details of decay products and their propagation, and thus widely 
applicable.

{\it Acknowledgments.---}
One of the author (K.I) would like to thank R. Nagata and K. Umezu for
useful discussions. K.T.'s work is supported by Grant-in-Aid for JSPS
Fellows, and K.I.'s work is supported in part by the Sasakawa Scientific
Research Grant from the JSS.

\end{document}